\documentclass[preprint,aps,amsmath]{revtex4}
\usepackage{graphicx}
\usepackage{bm}
\usepackage{amsmath}

\begin{document}
\preprint{}
\title{Turbulence and structure formation\\in complex plasmas and fluids}
\author{Alexander Kendl}
\affiliation{\small Institute for Ion Physics and Applied Physics, 
University of Innsbruck, A-6020 Innsbruck, Austria\vspace{2cm}}
\begin{abstract}
\vspace{0.5cm}
The formation and evolution of nonlinear and turbulent dynamical structures in
two-dimensional complex plasmas and fluids is explored by means of generalised
(drift) fluid simulations. Recent numerical results on turbulence in dusty
magnetised plasmas, strongly coupled fluids, semi-classical (``quantum'') plasmas and
in rotating quantum fluids are reviewed and discussed. 
\vspace{4cm}

{\sl \noindent This is the preprint version of:\\ 
AIP Conf. Proc. 1421 (2012), pp. 21-30; doi:http://dx.doi.org/10.1063/1.3679582 (10 pages):
INTERNATIONAL TOPICAL CONFERENCE ON PLASMA SCIENCE: Strongly Coupled
Ultra-Cold and Quantum Plasmas. Date: 12-14 September 2011. Location: Lisbon, Portugal.}
\end{abstract}
\maketitle

\section{INTRODUCTION}

Complex plasmas are characterised by strong coupling or complex interactions
among the constituing particles (electrons, one or more ion species, and
possibly neutral atoms and molecules), in particular in space and laboratory
plasmas which are cool, dense, and/or may contain a significant amount of
impurities, like massive dust particles \cite{Bonitz}.  

\bigskip

Strongly coupled plasmas are specifically characterized by a larger than unity
coupling coefficient, defined as the ratio between Coulomb energy and kinetic
energy, and for example occur in laser-produced plasmas or in compact
astrophysical objects, which in addition to strong coupling may for extreme
parameters also manifest quantum effects. Dusty plasmas occur in
interplanetary and interstellar space, in the edge of fusion experiments, and
are studied by dedicated laboratory experiments \cite{Fortov}. 
An other example for complex plasmas of topical interest are ultra-cold
neutral plasmas \cite{Killian07,Rolston08}, obtained in the  laboratory by
laser ionization of ultra-cold matter like Bose-Einstein condensates. 

\bigskip

The physics of complex plasmas has received considerable attention over recent
years. However, the theory and simulation of nonlinear dynamics and turbulent
structures in {\sl magnetised} complex plasmas so far remains rather unexplored -
despite the fact that static or dynamic magnetic fields are ubiquitously
present in many space and laboratory plasmas and lead to an abundance of
additional dynamical features. 

\bigskip

Then again, turbulence, vortices and flows in magnetised {\sl ideal} (high
temperature) plasmas have always been a topic of considerable interest for
fusion plasma physics in general, and in particular for the fusion related
previous work of the author of this contribution \cite{KendlIUP}.  
In this contribution we are going to explore the formation and evolution of
nonlinear and turbulent dynamical structures in magnetised complex plasmas by
means of generalised drift fluid simulations.

\bigskip

It should be stated clearly that hydrodynamic models for plasma turbulence, and
specifically in complex plasmas, most often can only be considered as a first
approximation to the problem. The importance of kinetic effects is rather a
rule than an exception, and for magnetised plasmas a 5-d gyrokinetic model can
in many cases be regarded as more appropriate, in particular if the
collisionality is low, particle trapping becomes important, or other kinetic
effects enter into the picture. If the relevant mode frequencies are not much
lower than the ion gyrofrequency then also the gyrokinetic model would have to be
replaced by a full Vlasov-Maxwell or N-particle model. These are in general
computationally very intensive and often not affordable for many realistic
problem scales. With noteable exceptions: recent ultra-cold neutral plasma
experients, for example, consist usually only of a number of particles in the
order of $10^6$ which is just manageable by advanced particle codes. 
For quantum condensate matter the numerical solution of the Gross-Pitaeveskii
equation (GPE) (based on the nonlinear Schroedinger equation) is standard and
delivers the full quantum physics involved.  
Such more realistic models, if available, should not be given up without need
on cost of simplified fluid models. 

\bigskip

Then again, simplified hydrodynamic models can be quite instructive in terms
of didactical aspects and may lend a more intuitive approach to the problem.
It is worthwile to investigate, in which minimal models certain effects still
qualitatively appear (even if the results may be quantitatively differing from
the fundamental models). And moreover, for some situations more fundamental
models than the fluid picture may even not be available (yet), or, as noted
above, too expensive to be solved numerically. The bottom line is that fluid
models may be instructive, but the user should be aware about their limitations.

\bigskip

Along this motivation, the Hasegawa-Wakatani (HW) model \cite{Hasegawa83} for
drift wave turbulence 
in magnetised plasmas is still often refered to for didactical purposes, even
since more sophisticates (gyrokinetic or gyrofluid) models have become generally available.  
HW is the minimal model allowing for fundamental insights into the drift
instability mechanism, nonlinear vortex development, and emergence of zonal
flows out of drift wave turbulence. 
The numerical solution of the 2-d HW equations is cheap and feasible even on a
contemporary laptop. We here want to use modified versions of the HW
equations for a first a proach to turbulence in complex magnetised plasmas.

\newpage
\section{2-DIMENSIONAL TURBULENCE MODEL}

The HW model for resistive drift wave turbulence \cite{Hasegawa83}
accounts for nonlinear instability driven by a gradient $\nabla n_0(x)$ in
plasma density and resistive parallel coupling between 
fluctuations of density $n$ and electrostatic potential $\phi$. 
The resulting turbulent state of ExB vortices in the $(x,y)$ drift plane
perpendicular to the magnetic field can form low-frequency $k_y=0$ zonal flow
structures with a finite wave number $k_x$.

The model is usually derived from isothermal electrostatic two-fluid equations
for electrons and ions. Under drift approximation the perpendicular momentum
equations deliver the low-frequency fluid drift velocities, notably the ExB
drift, diamagnetic drift and polarisation drift. In first order the ExB drift
velocity enters into the nonlinear advection term of the density (continuum)
equations, and the polarisation drifts enters through its finite divergence. 

Here we want to sketch out another approach and briefly motivate the derivation
of the HW equations from gyrokinetic theory. 
The gyrokinetic equation is an evolution equation for the 5-d distribution
function $f({\bf x}, v_{||}, v_{\perp}, t)$ with respect to guiding center
coordinates in a magnetised plasma. Modern gyrokinetic theory \cite{Brizard07}
is based on first stating the problem in Hamiltonian formalism and Lie transforming the
corresponding Euler-Lagrange equations to eliminate the gyroangle coordinate
under the assumption that the gyromotion is fast compared to all other
relevant time scales ($\omega \ll \Omega_i$).
The moment expansion into (gyro)fluid equations based on this procedure
involving a gyrokinetic Hamiltonian ideally conserves energy. 

Drift ordering further enters via smallness of the drift scale $\rho_s \ll
L_{\perp}$ compared to background gradient lengths and an ordering
$k_{\perp} \ll k_{||}$ with respect to orientation perpendicular and along the
magnetic field. A minimal form of a gyrokinetic equation may be constructed by
neglecting finite Larmor radius (FLR) effects, neglecting parallel dynamics,
assuming a homogeneous magnetic field, and keeping only ExB dynamics in the
convection, to obtain 
$\partial_t f +{\bf v}_{ExB} \cdot \nabla f = 0$.

Integration over velocity space is trivial when the most simple (fluid) model
for the distribution function $f= n F_0$ as a product between macroscopic
density $n$ and a Maxwellian $F_0$ is assumed. This results in advection
equations $\partial_t n_s +{\bf v}_{ExB} \cdot \nabla n_s = 0$ for the
densities $n_s$ of species $s=(e,i)$. These are coupled by the polarisation
equation $n_ee-n_ie = (n_0m_i/B^2)\nabla_{\perp}^2 \phi = \Omega$, relating the
densities to the vorticity $\Omega$. 

The ion density equation is usually replaced by a vorticity equation, which is obtained by
subtracting the density equations. For the parallel dynamics again the
simplest possible model is assumed, relating the electron current for cold
ions electrostatically to the density and potential via Ohm's law: $J_{||} =
(1/\eta_{||})[(T_e/n_ee)\nabla_{||}n_e-\nabla_{||}\phi]$.  

The resulting quasi-two-dimensional equations are normalised as 
\begin{equation}
\delta_0 e \phi/T_e \rightarrow \phi, \quad \delta_0 n_e/n_0 \rightarrow n+N(x), \quad
t c_s/L_{\perp} \rightarrow t, \quad x/L_{\perp} \rightarrow x, 
\end{equation}
with
$\delta_0 = \rho_s/L_{\perp}$, $c_s^2=T_e/m_i$, and $\rho_s^2=m_i T_e/(eB)^2$.
The density has been split into a fluctuating component $n$ and a constant
background $N(x)$. The dissipative coupling through the current is
parametrised by $D = k_{||}^2 ( L_{\perp}/c_s)/(T_e/n_o e^2 \eta_{||} )$.
The resulting set of equations is the HW standard model for resistive drift
wave turbulence: 

\begin{eqnarray}
 \partial_t n \; + \; [ \phi, n ] & = & - [\phi,N] + D ( \phi - n ) \label{hwden} \\
 \partial_t \Omega + [ \phi, \Omega ] & = & D ( \phi - n )   \label{hwvor} \\ 
\mbox{with} \quad \nabla_{\perp}^2 \phi & = & \Omega. \label{eq.poisson}
\end{eqnarray}

The general properties of the HW model have been extensively discussed
elsewhere (e.g. in  
Refs.~\cite{Koniges92,Biskamp94,Pedersen96,Zeiler96,Hu97,Camargo98,Korsholm03,Priego05,Numata07,Tynan07,Kendl11}).
The HW model supports unstable drift waves and saturated turbulence by tapping
the free energy in the background density gradient $N(x)$ through resisive
coupling via $D$. 

The hydrodynamic limit of the two-dimensional Euler equation is recovered for
$D=0$ and $N=const$, while the adiabatic limit $D \gg 1$ asymptotically
corresponds to the Hasegawa-Mima-Charney-Obukhov equation.
The Hasegawa-Mima (HM) system \cite{Hasegawa77} in itself is stable and may be
used to study decaying  turbulence, or it could be artificially driven. HM is
isomorphic to the Charney-Obukhov equations for rotating 2-d fluids (like
planetary atmospheres) including Rossby wave dynamics \cite{Horton94}. 

A characteristic property of these two-dimensional fluid systems is the
possibility for formation of large-scale zonal flow structures that are
coupled to the turbulent spectrum \cite{Diamond05}, which is a manifestation
of the dual cascade nature of 2-d turbulence.

The importance of this set of equations for plasma physics has been underlined
by the European Physical Society in awarding the 2011 Hannes Alfv\'en prize 
to Hasegawa, Mima and Diamond. Which is of course a bit ironic, as the
HW/HM models explicitly neglect Alfv\'en dynamics.

We numerically solve equations \eqref{hwden} and \eqref{hwvor} 
with an explicit 3rd  order Karniadakis time stepping
scheme \cite{Karniadakis}, and the Poisson brackets $[a,b] = (\partial_x
a)(\partial_y b) - (\partial_y a)(\partial_x b)$ are 
evaluated with the energy and enstrophy conserving Arakawa method
\cite{Arakawa}. The numerical method is equivalent to the one introduced in
Refs.~\cite{Naulin03,Scott05NJP}.  
Hyperviscuous operators $\nu^4\nabla^4$, with $\nu^4= -2\cdot10^{-4}$, are 
added for numerical stability to the right hand side of both equations
\eqref{hwden} and \eqref{hwvor}, acting on $n$ and $\Omega$, respectively.
We solve eq.~\eqref{eq.poisson} in $k$ space by evaluation of $\phi_k =
- \Omega_k / k_{\perp}^2$ employing the FFTW3 transform.  
The equations are discretised on a 2-d rectangular ($x$, $y$) grid with various
(in general not quadratic) box dimensions. Boundary conditions are periodic in
$y$ and either periodic or Dirichlet in $x$.
An example for a typical vorticity field $\Omega(x,y)$ in fully developed drift wave
turbulence is shown figure~1.

\bigskip

\begin{figure}[htb]
\centering
\includegraphics[width=16cm]{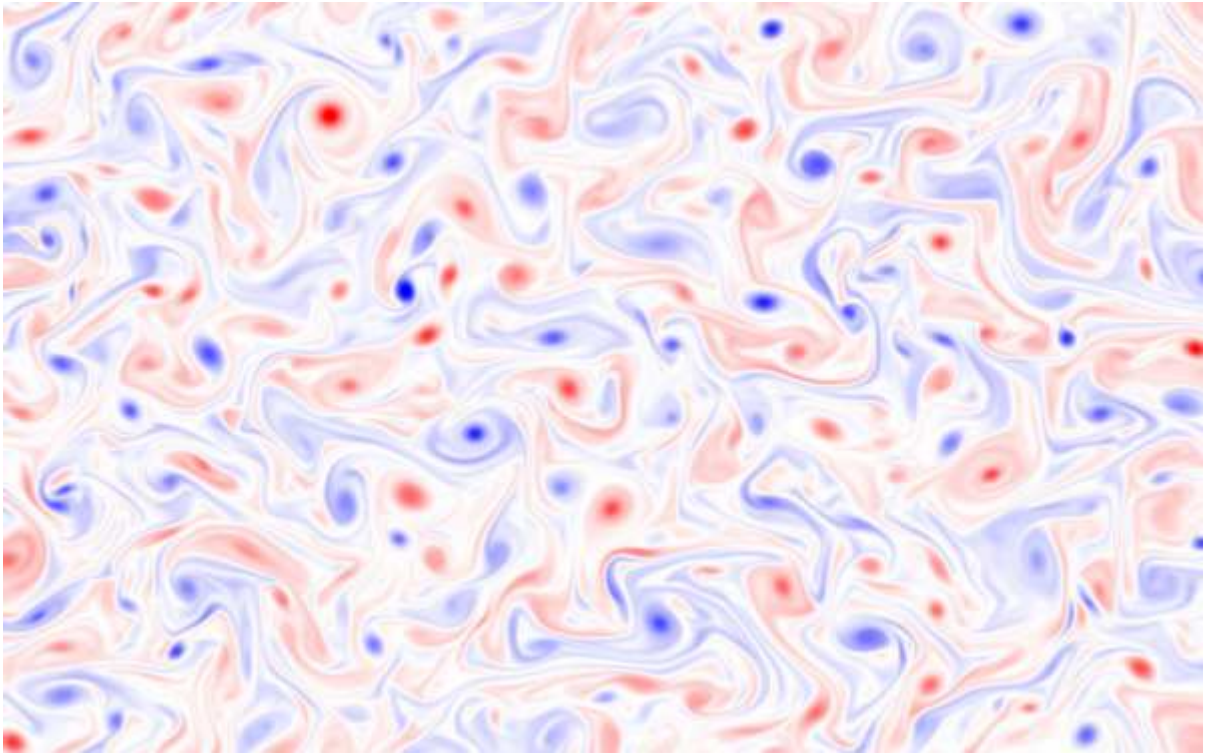}
\caption{\sl 
Vorticity $\Omega(x,y)$ in fully developed Hasegawa-Wakatani drift wave
turbulence. Negative values are colored in blue, positive values in red.
The physical domain is $64 \rho_s \times 128 \rho_s$ with a resolution of
$n_x=512$ and $n_y=1024$. The figure is here rotated by 90 degrees for better
display. The vortices can often be identified to appear in dipolar pairs.
 }
\end{figure}

\section{2-D TURBULENCE IN COMPLEX PLASMAS}

The motivation for the present contribution originated from discussions during
a visit of Padma K. Shukla to University of Innsbruck in 2010.
It became clear that a number of problems regarding waves and instabilities in
complex plasmas, that Shukla and many other authors had recently addressed by
means of linear theory or in 1-d, could often straightforwardly be
generalised to a 2-d nonlinear HW like system of fluid equations and put into a form
which is directly treatable by our numerical scheme. 
Shukla had in particular suggested to investigate turbulence in 2-d models for
quantum magnetoplasmas, dusty plasmas and strongly coupled fluids.
In the following we review the initial results of our simulations.

\subsection{Drift wave turbulence in the presence of a dust density gradient}

Shukla and Varma \cite{Shukla93} have proposed in 1993 a model to study the effect
of static, immobile dust grains on waves and instabilities in plasmas. It had
been shown that the presence of a static dust density gradient results in
modes with a frequency proportional to the dust density gradient scale.

This model has now been cast into a 2-d HW like form that allows the study of
drift wave turbulence in the presence of a density gradient of immobile charged
dust particles:
\begin{eqnarray}
\partial_t n \; + \; [ \phi, n ] & = & - (b+a) \; \partial_y \phi  + D ( \phi - n ),
 \label{dhwden} \\
\partial_t \Omega + [ \phi, \Omega ] & = & -a \; \partial_y \phi  
                                    - \gamma \; \Omega + D ( \phi - n),
\label{dhwvor} 
\end{eqnarray}
where $a =(\epsilon \rho_d Z_d c/B_0n_{i0}) \partial_x n_{d0}$ with
$\epsilon = (+1,-1)$ for (negative, positive) dust,
$\rho_d =c_d/\Omega_{ci}$, where $c_d = (\alpha k_B T_e/m_i)^{1/2}$ is the
 modified ion-sound speed \cite{Shukla92} with $\alpha =n_{i0}/n_{e0} > 1$.
The dust related viscosity is given by $\gamma$.
The constant plasma density gradient (derived from $N(x)$) is here parametrised by
$b=L_{\perp}/L_n$, which in our present normalisation is unity.
This set of modified HW equations has been derived and numerically solved by Shukla
and Kendl in ref.~\cite{Kendl11-dust}.
The quasi-linear drift wave instability through an ExB vortex growing out of
an initial Gaussian sensity perturbation has been studied for typical
resistive drift wave parameters and various dust gradient scales $a$.
It has been found that the presence of a co-aligned density gradient of
positively charged dust ($a>0$) strongly enhances the resistive drift wave
instability, while a counter-aligned dust gradient leads to a damping of the
drift waves. The same effect has been seen on fully developed turbulence:
co-aligned gradients result in larger fluctuation amplitudes and more
pronounced small-scale structures at high $k$, whereas counter-alignment damps
the turbulence into quasi-linear modes \cite{Kendl11-dust}.

The model so far has been restricted on immobile dust grains and did not
consider dust charging effects. A comprehensive 2-d multi-species gyrofluid
code for a more general modelling of dusty plasma turbulence is presently
being developed by the author.

\subsection{Generalised (viscoelastic) hydrodynamics}

Strongly coupled fluids are often modelled in ``generalised
hydrodynamics'' by including a viscoleastic relaxation time $\tau$ as
lowest order manifestation of kinetic coupling effects in
the (incompressible) Navier-Stokes equation: 
\begin{equation}
(1 + \tau \; \partial_t) \; d_t \; \Omega = \sigma \nabla_{\perp}^2 \Omega
\end{equation}
For numerical treatment with our existing solvers we split this equation in
two dimensions into the coupled set
\begin{eqnarray}
\partial_t \Omega + [ \phi, \Omega ] & = & -(1/\tau) (\Omega - \Psi) \\
\partial_t \Psi + [ \phi, \Omega ] & = &  \sigma \nabla_{\perp}^2 \Omega
\end{eqnarray}
so that for $\tau \rightarrow 0$ the Navier-Stokes equation is asymptotically
recovered. We have studied vortex stability, propagation and decay with this
system. Remarkably, we have found for comparable parameters qualitatively
nearly identical results as in a recent study \cite{Ashwin11} that employed a
``first principles'' molecular dynamics code to simulate coherent vortices in
strongly coupled liquids.

A personal favourite is the nonlinear evolution of a vortex with just the
right relation between initial amplitude and radius (i.e. half-width of a Gaussian).
It is well known from classical hydrodynamics that a strong vortex can develop
into a rotating tripole and further perform along various routes until
decay \cite{Fuentes96}. 

For certain parameters, the tripole can split up into
a dipole pair and a single monopolar vortex. For some parameters the single
may just hastily exit the system and leave the pair, which is further waltzing
around each other alone. 

For other initial parameters, the three can perform a
complicated kind of polka, where the separated monopole circles around the
rotating dipole in some distance, approaches the pair again to exchange
partners, and the freely relased vortex is now circling around the new pair... and
so on to viscosity.

A novel aspect regarding this dance, which we now studied with the above set of
equations, is the combined effect of viscosity $\sigma$ and viscoelastic relaxation
time $\tau$ on this triple polka. We find that by adding high viscosity
(e.g. $\sigma=0.1$) the initial vortex may rather (again depending on initial
amplitude and scale) 
disperse outwards and not further develop any nonlinear structures. Adding
finite $\tau$ (here in the range between 1 and 10) can compensate for the
dispersive action of $\sigma$, and the dipolar or tripolar vortex structures
of the inviscid case can be recovered.

\subsection{De Broglie screening effect in semi-classical plasmas}

Recently, Shukla and Kendl have derived and numerically solved a semi-classical
generalisation of the HW equations for dense degenerate Fermi plasmas including
quantum pressure corrections \cite{Kendl11-QHW}.
The model is based on the quantum magneto-hydrodynamic model \cite{Haas05,Brodin07}.
The dispersion relation of drift waves in magnetised quantum hydrodynamic plasmas
has before been studied by a number of authors in similar linear models,
e.g. refs. \cite{Shokri99,Ali07}. 

The semi-classical HW equations in an inhomogeneous magnetic field
are \cite{Kendl11-QHW}: 
\begin{eqnarray}
\partial_t \Omega + [ \phi, \Omega ] &=&
D ( \phi - \Lambda n )  - \kappa (\Lambda^{\!\ast} n) \label{qhw1} \\
\partial_t n \; + \; [ \phi, n ] &=& 
D ( \phi - \Lambda n )  + \kappa ( \phi - \Lambda^{\! \ast} n )
 \label{qhw2} 
\end{eqnarray}
where the curvature operator is defined as
\begin{equation}
\kappa (f) =  -  c \nabla \times \left( { {\bf B} / B^2 } \right) \cdot \nabla f  \nonumber  =  - c \nabla \times \left[({\bf B} \times \nabla f)/B^2\right].  
\end{equation}
The quantum corrections effect enters through 
$\Lambda = 1 - \beta^2\nabla^2$ and $\Lambda^{\! \ast} = 1 + \beta^2\nabla^2$,
with $\beta = \lambda_q/\rho_s$, where $\lambda_q = \hbar / \sqrt{4 m_e T_{F}}$
is the electron de Broglie length and $\rho_s$ the drift scale, both here defined
at Fermi temperature. 

The novel feature of this system is that we find a finite de Broglie
length (FBL) screening effect on density fluctuations, which is analogous to
the well-known FLR gyro screening and Debye screening of plasma turbulence.

An important caveat concerns the range of validity of the quantum hydrodynamic
model. The fluid-like equations including the quantum pressure are derived
from the nonlinear Schroedinger equations by means of an eikonal ansatz,
replacing the quantum mechanical wave function by an amplitude $n \sim
|\Psi|^2$, which follows a continuity equation, and a phase function $S$,
which dynamically evolves in space and time like a fluid velocity in Euler's
equation. It is important to remember that the eikonal ansatz is valid only in
the long wave length regime where all wave lengths $2\pi/k \gg \lambda_q$ are
required to be much larger than the de Broglie length. This is analogous to
the relation between electromagnetic wave optics and geometric ray optics. 
While this remark may perhaps seem obvious to most readers, it has to be noted
that some authors applying quantum plasma hydrodynamic models do
unfortunately not seem to be aware of these  restrictions. 

For quasi-linear drift waves the relevant wave lengths are in the order of a
few drift scales $\rho_s$, with most unstable modes usually found around
$0.1<\rho_s k<1$. Choosing the parameter $\beta = \lambda_q/\rho_s$ well below
unity is therefore a safe bet for linear theory. In fully developed turbulence
the cascade however in principle reaches down to viscous scales. We do not
have a good model of where to properly insert a viscous or Reynolds cut-off in dense
quantum plasma turbulence. Cutting the spectrum off by hyperviscosity at
scales related to $\rho_s/4$ seems tolerable in our experience. In any case we
require a small quantum parameter $\beta \ll 1$ for consistency. 

In our recent quantum effect studies on resistive drift wave
turbulence \cite{Kendl11-QHW} and interchange modes \cite{Kendl11-QIC} we have
used $\beta=0.5$ as a (perhaps already somewhat questionable) upper limit
which is barely consistent with the model validity.
The main message here is that our results are highly interesting from an
academic point of view, regarding the newly discovered FBL screening effect.
From a practical perspective, the quantum parameter in a valid range
$\beta \ll 1$  gives only minor corrections to the classical results. 
When compared to gyrofluids, this limit would somewhat correspond to a Taylor
approximation of the gyro-averaging operators -- which also is either practically
insignificant for most purposes if used in the correct limit, or otherwise
simply gives a wrong result. 

The validity of quantum hydrodynamics starts to cease at scales just where it
is beginning to become interesting. In particular, the important phenomenon of
vortex quantisation is not self-consistently treatable at all with the
hydrodynamic equations (other than being artifically modelled as specific point or
disc vortices).

\subsection{Quantum Charney-Obukhov turbulence}

The same notes of caution apply to the last quasi-2-d turbulent system, which
we here present and have recently addressed numerically. 
As mentioned in the introduction, there is a
well-known isomorphism between the Hasegawa-Mima and Charney-Obukhov
models. The latter describes rotating fluids in the presence of a gradient in
rotation frequency, which is applicable as a basic model for atmospherical
dynamics, and as solution includes Rossby waves and the secondary formation of
jet streams. Recent experiments on strongly rotating Bose-Einstein condensates
provided a motivation to apply the quantum hydrodynamic approach including
strong rotation. A quantum Charney-Obukhov equation has recently been
suggested and linearly analysed in ref.~\cite{Tercas10}:
\begin{equation}
\partial (\nabla^2 \psi - \phi) - \left[ \psi, \phi+\nabla^2 \psi - \ln
N \right] = 0
\end{equation}
where $\psi = \Lambda \phi$ with $\Lambda = 1 - \beta^2\nabla^2$.
This equation can be treated with our numerical scheme by explicitly evaluating 
$\partial W = [ \Lambda \phi, \phi+\nabla^2 \Lambda \phi - \ln N ]$, and
inverting $W_k = (1 + k^2 + (1/2) \beta^2 k^4)\phi_k$ in Fourier space.
We have numerically solved this system by using a Gaussian background rotation
$N(r)$ to model a rotating condensate in a trap and initially superimposing a
shear flow perturbation at half radius. The shear flow is found to be breaking
up into several vortex-like modes rotating around the center of the
condensate (see figure 2), and then is rather rapidly decaying, when the shear
flow perturbation is not continously driven. 

\bigskip

\begin{figure}[htb]
\centering
\includegraphics[width=6cm]{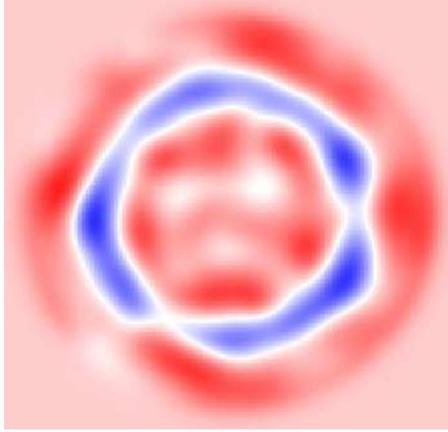}
\caption{\sl
A shear flow perturbation is superimposed on a circular rotating quantum fluid
in a 2-d numerical solution of the (actually semi-classical) quantum
Charney-Obukhov model. Here the 
perturbed stream function $\phi(x,y)$ is shown, which develops a modal structure
and (depending on parameters, here $\beta=0.8$) either rapidly decays or
transiently forms a chain of rotating vortices. Note the absence of vortex
quantisation in this simplified model. The color coding is blue for positive
(corresponding to the initial perturbation) and red for negative values of
$\phi$. Stream lines of the fluid flow are on isocontours of the stream
function. \vspace{1cm}
}
\end{figure}

To the best of our knowledge, such a nonlinear 2-d quantum
hydrodynamical model of rapidly rotating Bose-Einstein condensates has not
been simulated before. But again, in the valid range the small quantum factor
$\beta$ does not produce any significant effects. The behaviour of the system
is found to be very similar to a classical rotating fluid. 

What of course is completely absent in this model is vortex quantisation. And
as this is just the whole crux of quantum turbulence, the hydrodynamic
approach on BECs can in the personal opinion of the author be regarded as rather useless. 
Nothing new is learned, much is lost. There seems to be no good reason for not
rather solving the GPE (which is really rather standard now) to model BECs.

\section{CONCLUSIONS}

We have numerically studied the formation and evolution of nonlinear and
turbulent dynamical structures in two-dimensional complex plasmas and fluids.
Generalised (drift) fluid simulations have been developed on the basis of the
Hasegawa-Wakatani model and the 2-d Navier-Stokes vorticity equation.

Recently published results on turbulence in dusty magnetised plasmas and
semi-classical (``quantum'') plasmas have been reviewed. 

New results on 2-d generalised hydrodynamics including viscoelastic
relaxation effects were presented. The first nonlinear quantum
hydrodynamic simulation of a rotating Bose-Einstein condensate has been presented.
The limits of validity of the quantum hydrodynamic model with respect to
turbulence was critically discussed.

\section*{ACKNOWLEDGEMENTS}
The author thanks Padma K. Shukla for interesting discussions and for fruitful
cooperation on the cited joint publications. 
The work was supported by the Austrian Science Fund (FWF) Y398
and by a junior research grant from University of Innsbruck.

\end{document}